\documentclass[twocolumn,showpacs,aps,prl,superscriptaddress]{revtex4}

\usepackage{graphicx}
\usepackage{dcolumn}
\usepackage{amsmath}
\usepackage{epsfig}

\RequirePackage{xspace}

\usepackage{relsize}
\def\babar{\mbox{\slshape B\kern-0.1em{\smaller A}\kern-0.1em
    B\kern-0.1em{\smaller A\kern-0.2em R}}}

\def\epem       {\ensuremath{e^+e^-}\xspace}

\def\g     {\ensuremath{\gamma}\xspace}

\def\q     {\ensuremath{q}\xspace}

\def\qqbar {\ensuremath{q\overline q}\xspace}

\def\pip   {\ensuremath{\pi^+}\xspace}
\def\pim   {\ensuremath{\pi^-}\xspace}

\def\pipm  {\ensuremath{\pi^\pm}\xspace}
\def\pimp  {\ensuremath{\pi^\mp}\xspace}

\def\Kbar  {\kern 0.2em\overline{\kern -0.2em K}{}\xspace}

\def\Kz    {\ensuremath{K^0}\xspace}
\def\Kzb   {\ensuremath{\Kbar^0}\xspace}
\def\KzKzb {\ensuremath{\Kz \kern -0.16em \Kzb}\xspace}
\def\Kp    {\ensuremath{K^+}\xspace}
\def\Km    {\ensuremath{K^-}\xspace}
\def\Kpm   {\ensuremath{K^\pm}\xspace}
\def\Kmp   {\ensuremath{K^\mp}\xspace}
\def\KpKm  {\ensuremath{\Kp \kern -0.16em \Km}\xspace}
\def\KS    {\ensuremath{K^0_{\scriptscriptstyle S}}\xspace} 
 
\def\Kstarz  {\ensuremath{K^{*0}}\xspace}
\def\Kstarzb {\ensuremath{\Kbar^{*0}}\xspace}
\def\Kstar   {\ensuremath{K^*}\xspace}

\def\Kstarpm {\ensuremath{K^{*\pm}}\xspace}

\newcommand{\etapr}{\ensuremath{\eta^{\prime}}\xspace}

\def\Dbar    {\kern 0.2em\overline{\kern -0.2em D}{}\xspace}

\def\Dz      {\ensuremath{D^0}\xspace}
\def\Dzb     {\ensuremath{\Dbar^0}\xspace}
\def\DzDzb   {\ensuremath{\Dz {\kern -0.16em \Dzb}}\xspace}
\def\Dp      {\ensuremath{D^+}\xspace}
\def\Dm      {\ensuremath{D^-}\xspace}

\def\DpDm    {\ensuremath{\Dp {\kern -0.16em \Dm}}\xspace}

\def\Ds      {\ensuremath{D^+_s}\xspace}

\def\B       {\ensuremath{B}\xspace}
\def\Bbar    {\kern 0.18em\overline{\kern -0.18em B}{}\xspace}

\def\BB      {\ensuremath{B\Bbar}\xspace} 
\def\Bz      {\ensuremath{B^0}\xspace}
\def\Bzb     {\ensuremath{\Bbar^0}\xspace}
\def\BzBzb   {\ensuremath{\Bz {\kern -0.16em \Bzb}}\xspace}
\def\Bu      {\ensuremath{B^+}\xspace}
\def\Bub     {\ensuremath{B^-}\xspace}
\def\Bp      {\ensuremath{\Bu}\xspace}
\def\Bm      {\ensuremath{\Bub}\xspace}

\def\BpBm    {\ensuremath{\Bu {\kern -0.16em \Bub}}\xspace}

\def\BorBbar    {\kern 0.18em\optbar{\kern -0.18em B}{}\xspace}
\def\DorDbar    {\kern 0.18em\optbar{\kern -0.18em D}{}\xspace}
\def\KorKbar    {\kern 0.18em\optbar{\kern -0.18em K}{}\xspace}

\def\jpsi     {\ensuremath{{J\mskip -3mu/\mskip -2mu\psi\mskip 2mu}}\xspace}
\def\psitwos  {\ensuremath{\psi{(2S)}}\xspace}

\mathchardef\Upsilon="7107
\def\Y#1S{\ensuremath{\Upsilon{(#1S)}}\xspace}

\def\FourS {\Y4S}

\mathchardef\Deltares="7101
\mathchardef\Xi="7104
\mathchardef\Lambda="7103
\mathchardef\Sigma="7106
\mathchardef\Omega="710A

\def\Deltabar{\kern 0.25em\overline{\kern -0.25em \Deltares}{}\xspace}
\def\Lbar{\kern 0.2em\overline{\kern -0.2em\Lambda\kern 0.05em}\kern-0.05em{}\xspace}
\def\Sigbar{\kern 0.2em\overline{\kern -0.2em \Sigma}{}\xspace}
\def\Xibar{\kern 0.2em\overline{\kern -0.2em \Xi}{}\xspace}
\def\Obar{\kern 0.2em\overline{\kern -0.2em \Omega}{}\xspace}
\def\Nbar{\kern 0.2em\overline{\kern -0.2em N}{}\xspace}
\def\Xb{\kern 0.2em\overline{\kern -0.2em X}{}\xspace}

\def\mes        {\mbox{$m_{\rm ES}$}\xspace}

\def\DeltaE     {\mbox{$\Delta E$}\xspace}

\newcommand{\tev}{\ensuremath{\mathrm{\,Te\kern -0.1em V}}\xspace}
\newcommand{\gev}{\ensuremath{\mathrm{\,Ge\kern -0.1em V}}\xspace}
\newcommand{\mev}{\ensuremath{\mathrm{\,Me\kern -0.1em V}}\xspace}
\newcommand{\kev}{\ensuremath{\mathrm{\,ke\kern -0.1em V}}\xspace}
\newcommand{\ev}{\ensuremath{\mathrm{\,e\kern -0.1em V}}\xspace}
\newcommand{\gevc}{\ensuremath{{\mathrm{\,Ge\kern -0.1em V\!/}c}}\xspace}
\newcommand{\mevc}{\ensuremath{{\mathrm{\,Me\kern -0.1em V\!/}c}}\xspace}
\newcommand{\gevcc}{\ensuremath{{\mathrm{\,Ge\kern -0.1em V\!/}c^2}}\xspace}
\newcommand{\mevcc}{\ensuremath{{\mathrm{\,Me\kern -0.1em V\!/}c^2}}\xspace}

\def\invfb   {\ensuremath{\mbox{\,fb}^{-1}}\xspace}

\def\mus  {\ensuremath{\rm \,\mus}\xspace}

\def\mus        {\ensuremath{\,\mu{\rm s}}\xspace}

\def\to                 {\ensuremath{\rightarrow}\xspace}

\def\pep2{PEP-II}

\def\gsim{{~\raise.15em\hbox{$>$}\kern-.85em
          \lower.35em\hbox{$\sim$}~}\xspace}
\def\lsim{{~\raise.15em\hbox{$<$}\kern-.85em
          \lower.35em\hbox{$\sim$}~}\xspace}

\xspace

\def\jetset74   {\mbox{\tt Jetset \hspace{-0.5em}7.\hspace{-0.2em}4}\xspace}

\newcommand{\splot}    {\mbox{$_s{\cal P}lot$}\xspace}

\newcommand{\onreslumi}  {\mbox{424\invfb}}
\newcommand{\offreslumi} {\mbox{44\invfb}}
\newcommand{\bbpairs}    {\mbox{$(465\pm5)\times 10^6$}}

\def\ncand    {\ensuremath{14\,276}}
\def\nsig     {\ensuremath{262\pm 47}}
\def\nBBtwo   {\ensuremath{199\pm 51}}
\def\nBBthree {\ensuremath{262\pm 55}}
\def\nExpBBtwo  {\ensuremath{173}}
\def\nExpBBthree{\ensuremath{215}}
\def\nsigEffCor{\ensuremath{1326\pm 207}}
\def\nsigma   {\ensuremath{6.0\,\sigma}}
\def\nsigSys  {\ensuremath{5.2\,\sigma}}
\def\kskpiBFal{\ensuremath{\left(3.2\pm 0.5\pm 0.3\right)\times 10^{-6}}}

\newcommand{\BABARPubYear}    {09}
\newcommand{\BABARPubNumber}  {036}
\newcommand{\SLACPubNumber} {14001}

\begin{document}

\preprint{\babar-PUB-\BABARPubYear/\BABARPubNumber} 
\preprint{SLAC-PUB-\SLACPubNumber} 

\begin{flushleft}
\babar-PUB-\BABARPubYear/\BABARPubNumber\\
SLAC-PUB-\SLACPubNumber\\
%\LANLNumber\\[10mm]
\end{flushleft}

\title{
  {\large 
    \bf Observation of the Rare Decay {\boldmath $\Bz\to\KS\Kpm\pimp$}
  }
}

\author{P.~del~Amo~Sanchez}
\author{J.~P.~Lees}
\author{V.~Poireau}
\author{E.~Prencipe}
\author{V.~Tisserand}
\affiliation{Laboratoire d'Annecy-le-Vieux de Physique des Particules (LAPP), Universit\'e de Savoie, CNRS/IN2P3,  F-74941 Annecy-Le-Vieux, France}
\author{J.~Garra~Tico}
\author{E.~Grauges}
\affiliation{Universitat de Barcelona, Facultat de Fisica, Departament ECM, E-08028 Barcelona, Spain }
\author{M.~Martinelli$^{ab}$}
\author{A.~Palano$^{ab}$ }
\author{M.~Pappagallo$^{ab}$ }
\affiliation{INFN Sezione di Bari$^{a}$; Dipartimento di Fisica, Universit\`a di Bari$^{b}$, I-70126 Bari, Italy }
% ...
\collaboration{The \babar\ Collaboration}
\noaffiliation

\date{\today}

\begin{abstract}
We report an analysis of charmless
hadronic decays of neutral \B\ mesons to the final state $\KS\Kpm\pimp$,
using a data sample of \bbpairs\ \BB\ events collected with the
\babar\ detector at the \FourS\ resonance.
We observe an excess of signal events with a significance of 
$5.2$ standard deviations including systematic uncertainties and measure
the branching fraction to be ${\cal B}\left(\Bz\to\KS\Kpm\pimp\right) =
\kskpiBFal$, where the uncertainties are statistical and systematic,
respectively. 

\end{abstract}


\maketitle

Charmless decays of \B\ mesons to hadronic final states containing an even
number of kaons are suppressed in the standard model (SM). 
Decays of this type mainly proceed via the $b\to d$ ``penguin'' transition,
involving a virtual loop, and hence are sensitive to potential new physics
contributions since the presence of new particles in the loops can
produce deviations from SM expectations.
In recent years, there has been a surge of new results on these decays:
$\Bz\to\KS\KS$ and $\Bp\to\KS\Kp$ have been observed~\cite{Aubert:2006gm,Abe:2006xs},
and there is evidence for the related vector-vector
final states~\cite{Aubert:2007xc,Aubert:2009vc,Chiang:2010ga}.
Only upper limits on the corresponding pseudoscalar-vector final states exist:
${\cal B}(\Bz\to\Kz\Kstarzb) + 
{\cal B}(\Bz\to\Kzb\Kstarz) < 1.9 \times 10^{-6}$~\cite{Aubert:2006wu}
and
${\cal B}(\Bp\to\Kp\Kstarzb) < 1.1 \times 10^{-6}$~\cite{Aubert:2007ua},
both at 90\,\% confidence level
(unless explicitly stated otherwise we use the symbol \Kstar\ to denote the
$\Kstar(892)$ resonance and the inclusion of charge conjugate modes is implied).
Note that decays with additional suppression in the SM, 
such as $\Bz\to K^{(*)+}K^{(*)-}$, which are expected to proceed via annihilation amplitudes,
have not been observed~\cite{Aubert:2006fha,Abe:2006xs,Aaltonen:2008hg,:2008ap,Aubert:2007xc,Aubert:2007fm,Aubert:2008rr,Aubert:2006aw}.

Since the vector resonances involved have non-negligible widths,
the pseudoscalar-vector decays are best studied using Dalitz plots of the
three-body $KK\pi$ final states. In the three-body channels, contributions from
suppressed $b\to u$ tree amplitudes are expected to be important, in
addition to the $b\to d$ penguin amplitudes. Recent investigations of
three-body channels suggest that additional resonances are present.
Most notably, the $\Bp\to\Kp\Km\pip$ channel exhibits an unexpected peak near
$1.5\gevcc$ in the $\Kp\Km$ invariant-mass spectrum, which accounts for approximately
half of the total event rate~\cite{Aubert:2007xb}.
We call this peak, with unknown spin and isospin quantum 
numbers, the $f_{\rm X}(1500)$.
The lack of a $f_{\rm X}(1500)$ signal
in $\Bu\to\KS\KS\pip$ decays implies that the $f_{\rm X}(1500)$ does
not have even spin if isospin is conserved in the decay~\cite{Aubert:2008aw}.
A search for an isospin partner to the $f_{\rm X}(1500)$ that decays to
$\Kzb\Kp$ and which could be produced recoiling against a pion
in \B\ decay could help to clarify the nature of this resonance.

In this paper, we present the results of a search for the three-body decay
$\Bz\to\KS\Kpm\pimp$, including intermediate two-body modes that decay to this
final state but do not contain charm quarks.
No decays to this final state have been observed as yet.
The best available upper limit on the inclusive branching fraction is 
${\cal B}(\Bz\to\Kz\Kpm\pimp)<18\times 10^{-6}$~\cite{Garmash:2003er}.
There appears to be no explicit prediction for the inclusive branching fraction
of $\Bz\to\KS\Kpm\pimp$.
Some theoretical predictions exist, however, for the relevant resonant modes. 
Expected branching fractions for $\Bz\to(\Kstarzb\Kz + \Kstarz\Kzb)$ and 
$\Bz\to\Kstarpm\Kmp$ are in the range $(0.2$--$2.0)\times 10^{-6}$ and
$(0.2$--$1.0)\times 10^{-7}$, respectively~\cite{Du:1995ff,Ali:1998eb,Chen:1999nxa,Deshpande:1997rr,Du:2002up,Beneke:2003zv,Chiang:2003pm,Guo:2006uq}.
Extensions to the SM can yield significantly larger branching fractions. 
For instance, in supersymmetric models with $R$-parity violation, the branching
fraction for $\Bz\to(\Kstarzb\Kz + \Kstarz\Kzb)$ could be as large as 
$10^{-5}$~\cite{Wang:2006sy}. 

The data used in the analysis, collected with the
\babar\ detector~\cite{Aubert:2001tu} at the \pep2\ asymmetric energy 
\epem\ collider at SLAC, consist of an integrated luminosity of \onreslumi\ recorded
at the \FourS\ resonance (``on-peak'') and \offreslumi\
collected 40\,\mev\ below the resonance (``off-peak'').
The on-peak data sample contains \bbpairs\ \BB\ events.

We reconstruct $\Bz\to\KS\Kpm\pimp$ decay candidates by 
combining a \KS\ candidate with one charged kaon
and one oppositely charged pion candidate. 
The $\Kpm$ and $\pipm$ candidates are required to have a minimum
transverse momentum of 50\,\mevc\ and to be consistent with having originated
from the interaction region. Identification of charged kaons and pions is
accomplished with energy-loss information from the tracking subdetectors, and
the Cherenkov angle and number of photons measured by a
ring-imaging Cherenkov detector. We distinguish kaons from pions by applying 
criteria to the product of the likelihood ratios determined from these
individual measurements. The efficiency for kaon selection is approximately
80\,\% including geometrical acceptance, while the probability of
misidentification of pions as kaons is below 5\,\% up to a laboratory momentum
of 4\,\gevc. 
A $\KS\to\pip\pim$ candidate is formed from a pair of oppositely charged
tracks (with the pion mass hypothesis assumed) having an invariant mass that lies
within 15\mevcc\ of the nominal \KS\ mass~\cite{Amsler:2008zz}, 
corresponding to $5$ times the \KS\ mass resolution.
We require the ratio of the measured \KS\ decay length and its uncertainty to be
greater than 20, the cosine of the angle between the line connecting the 
\B\ and \KS\ decay vertices and the \KS\ momentum vector to be greater than
0.999, and the \KS\ vertex fit probability to be greater than $10^{-6}$.

To suppress the dominant background contribution, which arises from continuum
$\epem\to\qqbar\ (\q=u,d,s,c)$ events, we employ a Fisher discriminant that
combines four variables. These are the ratio of the second to the zeroth order
momentum-weighted angular moment~\cite{Aubert:2009az}, the absolute
value of the cosine of the angle between the \B\ direction and the beam
axis, the magnitude of the cosine of the angle between the \B\ thrust axis
and the beam axis, and the proper time difference between
the decays of the two \B\ mesons divided by its statistical uncertainty. 
The first three quantities are calculated in the center-of-mass (CM) frame.

In addition to the Fisher output (${\cal F}$), we distinguish signal from
background events using two kinematic variables: the difference \DeltaE\
between the CM energy of the \B\ candidate and $\sqrt{s}/2$, and the
beam-energy-substituted mass $\mes=\sqrt{s/4-{\bf p}^2_\B}$, where $\sqrt{s}$
is the total CM energy and ${\bf p}_\B$ is the momentum of the candidate \B\
meson in the CM frame.
The signal \mes\ distribution peaks near the \B\ mass with a resolution of
about $2.6\mevcc$, while its \DeltaE\ distribution peaks at zero with a
resolution of approximately $20\mev$.
We select signal candidates that satisfy $5.272<\mes<5.286\gevcc$,
$|\DeltaE|<0.075\gev$, and ${\cal F}>-0.145$.
The requirement on ${\cal F}$ removes approximately 70\,\% of continuum
background while retaining 90\,\% of signal events.

Another source of background arises from \B\ decays, mostly involving
intermediate charm or charmonium mesons, or charmless final states that are
misreconstructed. We exclude \B\ candidates that have two-body mass
combinations in any of the following invariant-mass ranges: 
$1.82 < m(\KS\Kpm) < 2.04$, $1.81 < m(\KS\pimp) < 1.91$, 
$1.83 < m(\Kpm\pimp) < 1.90$, $3.06 < m(\Kpm\pimp) < 3.17$,
and $3.66 < m(\Kpm\pimp) < 3.73$ (all in units of \gevcc).
These ranges reject decays from \Dp\ and \Ds, \Dp, \Dz, \jpsi, and \psitwos
mesons, respectively. Charmonium contributions result mainly from the leptonic
decays of \jpsi\ and \psitwos, where one lepton is misidentified as a
charged pion and the other as a kaon.

The efficiency for signal events to pass all the selection criteria is
determined as a function of position in the Dalitz plot.
Using a Monte Carlo (MC) simulation in which events uniformly populate the
phase-space, we obtain an average efficiency of 20\,\%, though
values as high as double (as low as half) that value are found near the
center (corners) of the Dalitz plot.

An average of $1.1$ \B\ candidates is found per selected event.
In events with multiple candidates we choose the one with the highest
\B\ vertex fit probability. 
We verify that this procedure does not bias our fit variables.
In some signal events, the \B\ candidate is misreconstructed due to
one track being replaced with a track from the rest of the event. 
The fraction of such events is below 2\,\% in the phase-space MC, but is
closer to 5\,\% in MC samples where the events populate the $\Kstar$ bands.
Misreconstructed signal events are considered as a part of the signal
component in the fit described below.
We assign a systematic error to account for the uncertainty in the rate of
these events, which is related to the unknown Dalitz-plot distribution of the
$\Bz\to\KS\Kpm\pimp$ decay.

We study residual background contributions from \BB\ events that survive
the invariant-mass exclusion requirements described earlier, using MC
simulations. It is found that these events can be combined
into four categories based on their shapes in \mes\ and \DeltaE.
The first category ($\BB_1$) comprises $\Bz\to\etapr\KS,\etapr\to\rho^0\g$
and misreconstructed $\Bz\to\Dm\pip,\Dm\to\KS\Km$ decays and has a broad peak
in \mes\ and a nonpeaking \DeltaE\ shape.
The second and third categories ($\BB_2$ and $\BB_3$) represent the charmless
decays $\Bz\to\KS\Kp\Km$ and $\Bz\to\KS\pip\pim$, where a kaon or a pion is
misidentified leading to a \DeltaE\ distribution that peaks with negative or
positive mean, respectively. 
The MC simulations of these decays are based on our recent studies of
their Dalitz plot distributions~\cite{Aubert:2007sd,Aubert:2007vi}.
The fourth category ($\BB_4$) contains the remainder of the \BB\ background
and is mainly combinatorial in nature. 
Based on the MC-derived efficiencies, total number of \BB\ events, and known
branching fractions~\cite{hfag,Amsler:2008zz}, we expect $25$, \nExpBBtwo, 
\nExpBBthree, and $668$ events from the four \BB\ background categories,
respectively. 

To obtain the $\Bz\to\KS\Kpm\pimp$ signal yield, we perform an
unbinned extended maximum likelihood fit to the candidate events using
three input variables: \mes, \DeltaE, and ${\cal F}$. These
variables are found to be largely uncorrelated -- the maximum correlation is
between the signal \mes\ and \DeltaE\ distributions and is about 13\,\%. 
For each component $j$ (signal, \qqbar\ background, and the four 
\BB\ background categories), we define a probability density function (PDF) 
\begin{equation}
  \label{eq:PDF-exp}
  {\cal P}^i_j \equiv
  {\cal P}_j(\mes^i) {\cal P}_j(\DeltaE^i) {\cal P}_j({\cal F}^i),
\end{equation}
where $i$ denotes the event index. The extended likelihood
function is given as
\begin{equation}
  \label{eq:extML-Eq}
  {\cal L} =
  \prod_{k} e^{-n_k}
  \prod_{i}\left[ \sum_{j}n_j{\cal P}^i_j \right],
\end{equation}
where $n_{j(k)}$ is the yield of the event category $j(k)$.

For the signal component, the \mes\ and \DeltaE\ distributions are
each parametrized by the sum of two Gaussian functions, 
while the ${\cal F}$ distribution
is described by a bifurcated Gaussian function with a small admixture from the
sum of two Gaussians. 
We fix the shape parameters to the
values obtained from the $\Bz\to\KS\Kpm\pimp$ phase-space MC sample,
after adjusting them to account for possible differences between data
and MC simulations determined with a control sample of $\Bz\to\Dm\pip,
\Dm\to\KS\pim$ decays. For the continuum background, we use an ARGUS
function~\cite{Albrecht:1990am} to parametrize the \mes\ shape and a
linear function for \DeltaE. The continuum Fisher shape is modeled
with a function that is composed of a Gaussian tail with
relative fraction 99.6\,\% (large component) and a small Gaussian
with different mean and width values.
This shape provides a good description of the off-peak Fisher
distribution, as well as of the corresponding MC distribution.
One-dimensional histograms are used as nonparametric
PDFs to represent all three fit variables for the four \BB\ background
components.

The free parameters of our fit are the yields of signal, $\BB_2$,
$\BB_3$, and continuum background together with the slope of the continuum
$\DeltaE$ PDF and the mean and width of the large Gaussian component of the
continuum ${\cal F}$ PDF. 
The ARGUS $\xi$ parameter and parameters of the small Gaussian
component of the continuum Fisher function are fixed to values determined
from candidates selected in the off-peak data sample with a looser
requirement on \mes. The yields of $\BB_1$ and $\BB_4$, and
all shape parameters of the four \BB\ background categories are
fixed to the values determined from MC simulations.

We cross check our analysis procedure by removing the requirements that reject
backgrounds from \B\ decays involving charm mesons, 
instead selecting regions of the Dalitz plot dominated by intermediate charm
mesons. 
We select $\Bz\to\Dm\pip,\Dm\to\KS\Km$ decays requiring
$1.84<m(\KS\Kpm)<1.89\gevcc$, and $\Bz\to\Dm\Kp,\Dm\to\KS\pim$ decays
requiring $1.84<m(\KS\pipm)<1.89\gevcc$. We then apply our fit to find
the yields for the $\Bz\to\Dm\pip$ and $\Bz\to\Dm\Kp$ channels.
We find values consistent with the expectations
based on world-average product branching fractions~\cite{Amsler:2008zz}
within statistical uncertainties.

\begin{figure*}[!htb]
\center
\includegraphics[width=0.35\textwidth]{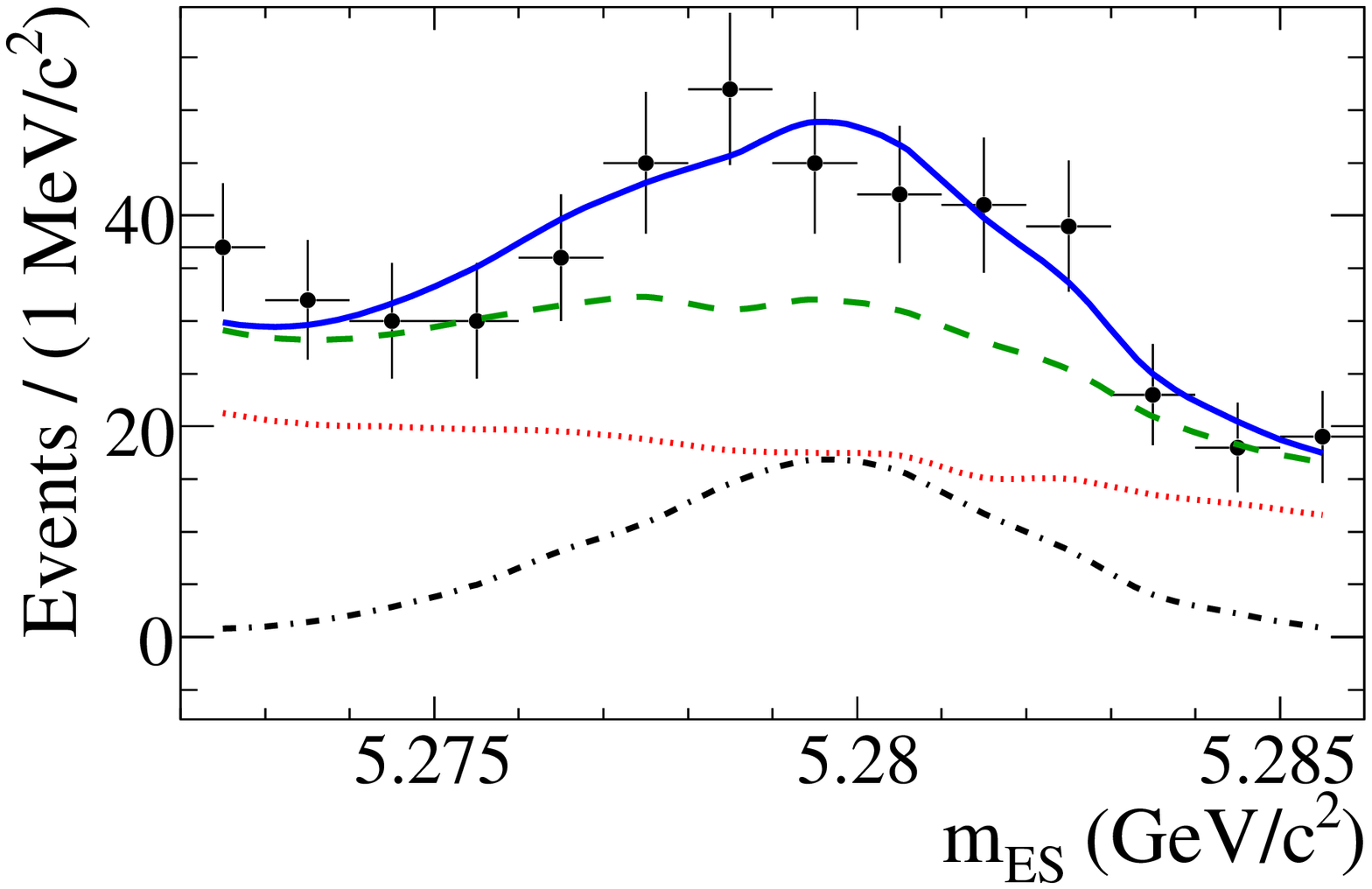}
\includegraphics[width=0.35\textwidth]{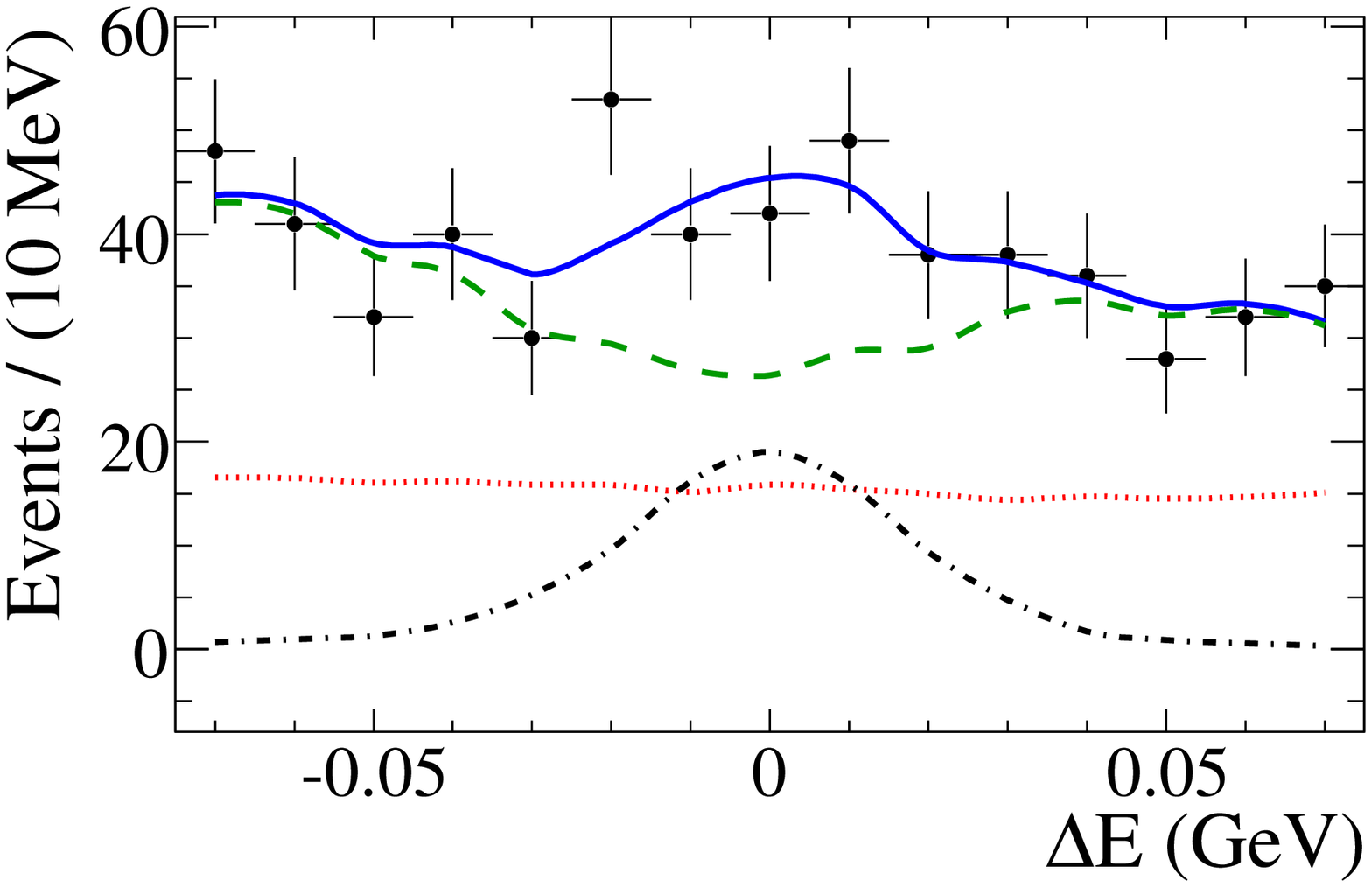} \\
\includegraphics[width=0.35\textwidth,bb=10 0 555 360,clip=true]{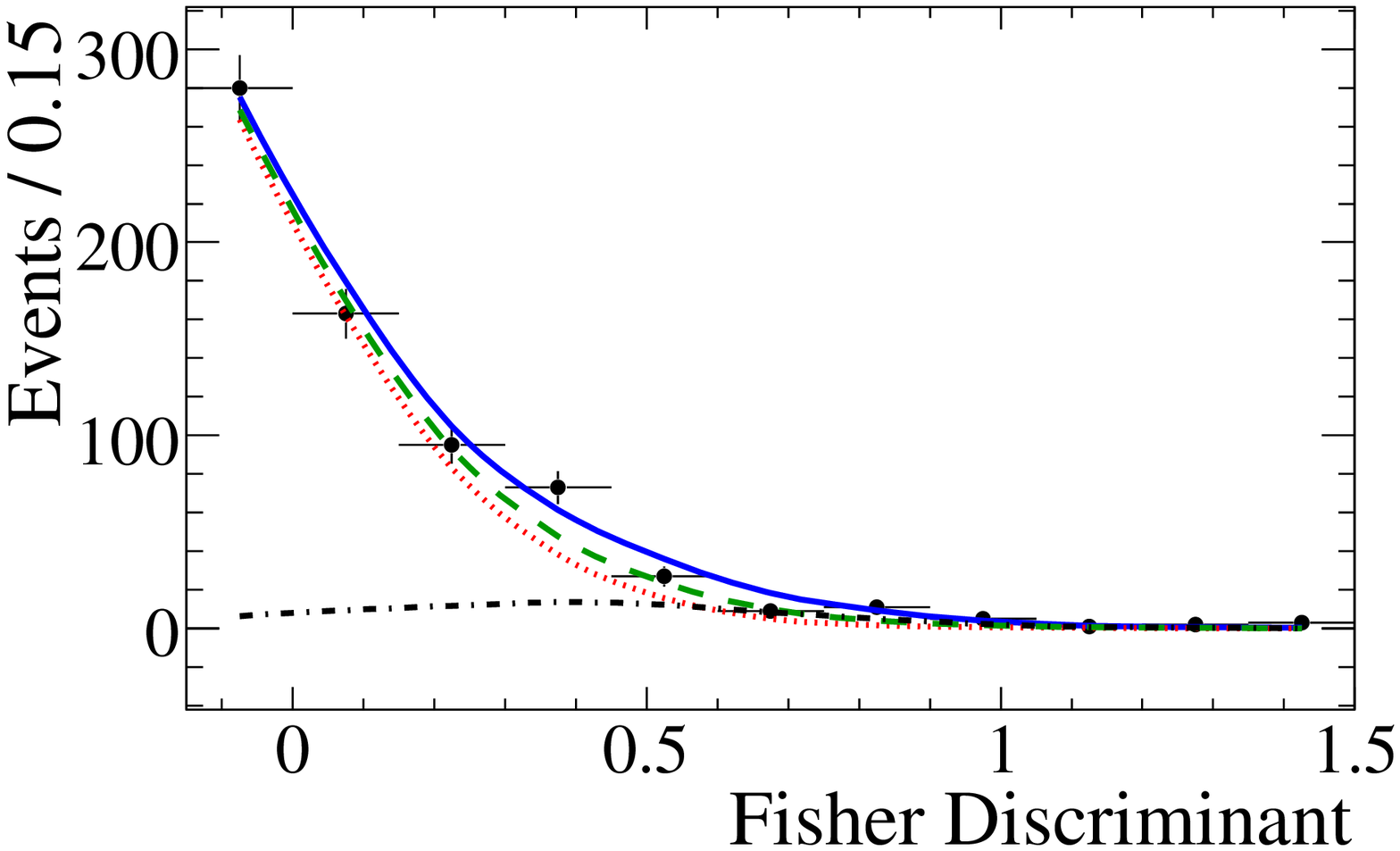}
\includegraphics[width=0.35\textwidth,bb=0 0 555 355,clip=true]{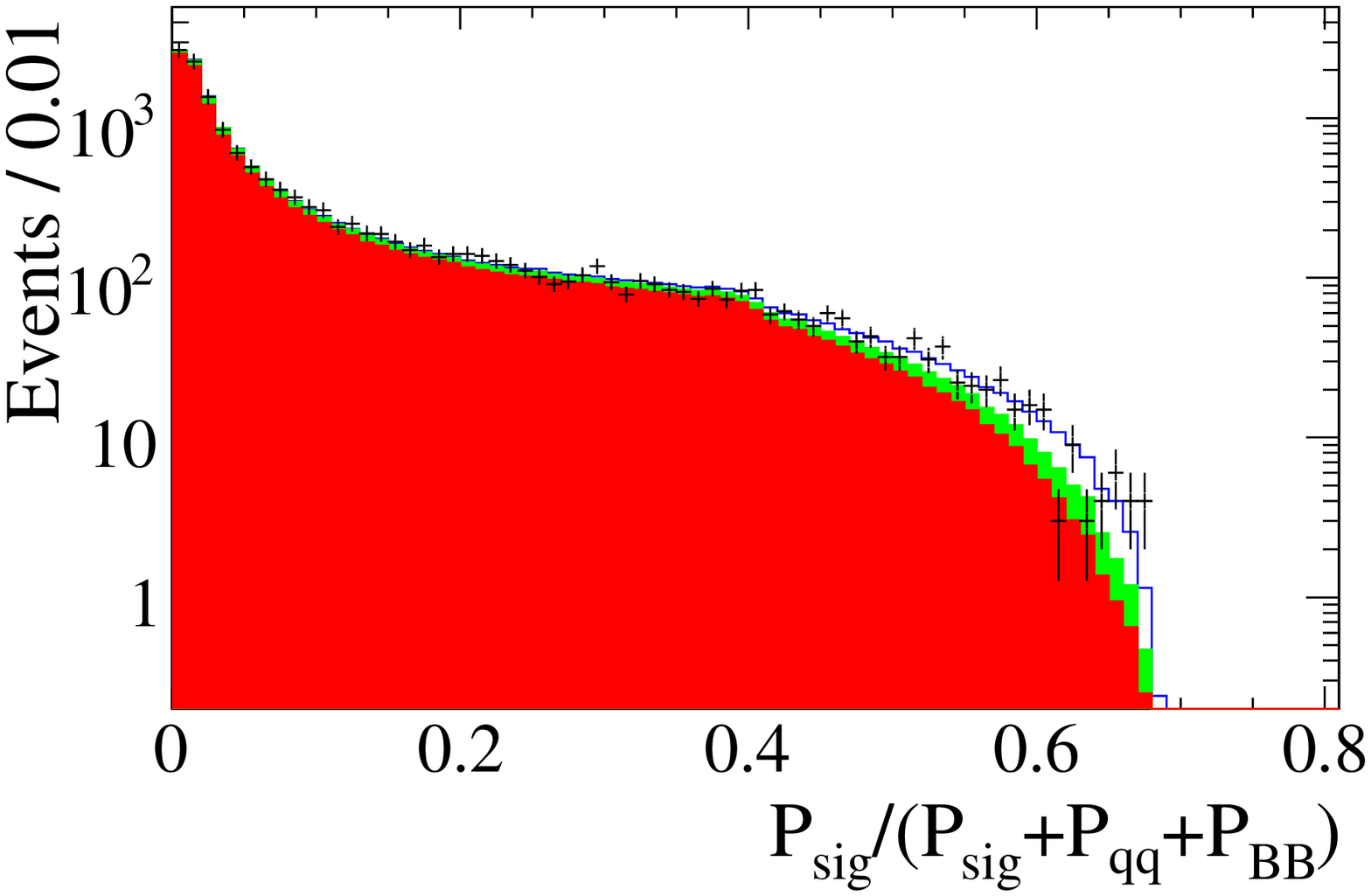}
\caption{(color online).
  Projections of candidate events onto
  (top left) \mes, (top right) \DeltaE\ and (bottom left) ${\cal F}$,
  following a requirement on the likelihood ratio to enhance signal
  visibility, calculated without the plotted variable. 
  Points with error bars show the data, the solid (blue)
  curves are the total fit result, the dotted (red) curves are the \qqbar
  component, and the dashed (green) curves show the total background
  contribution. The dash-dotted curves represent the signal contribution.
  Note that the $\BB_2$ and $\BB_3$ background categories have 
  \DeltaE\ distributions that peak with negative and positive mean values,
  respectively.
  The bottom right plot shows the distribution of the likelihood ratio
  ${\cal P}_{\rm sig}/({\cal P}_{\rm sig}+{\cal P}_{\BB}+{\cal P}_{\qqbar})$
  for all candidate events.
  Points with error bars show the data, 
  the dark (red) filled area shows the contribution from \qqbar\ background, 
  the light (green) filled area shows the contribution from \BB\ background,
  and the solid (blue) histogram shows the sum of all contributions.
}
\label{fig:signal-project}
\end{figure*}

Applying the fit method described above to the \ncand\ selected candidate
$\Bz\to\KS\Kpm\pimp$ events, we find \nsig\ signal events.
The fitted yields of the $\BB_2$ and $\BB_3$ categories are \nBBtwo\ and
\nBBthree, respectively, consistent with the MC-based expectations.
The fitted values of all other free parameters of the fit are also consistent
with expectations based on studies of control samples and MC simulations.
The results of the fit are shown in Fig.~\ref{fig:signal-project}.
The statistical significance of the signal yield, 
given by the square root of the difference between twice the value of 
negative log likelihood obtained assuming zero signal events to that at its
minimum, is \nsigma.
Including systematic uncertainties (discussed below), the significance is
\nsigSys.

The $\Bz\to\KS\Kpm\pimp$ branching fraction is determined from the result of
the fit by calculating signal probabilities for each candidate event with the
\splot\ technique~\cite{Pivk:2004ty}. 
These are divided by event-by-event efficiencies that take the Dalitz-plot
position dependence into account, and summed to obtain an efficiency-corrected
signal yield of \nsigEffCor\ events.
We further correct for the effect of the charm and charmonium vetoes
(estimated using a range of MC samples with different Dalitz-plot
distributions), 
and divide by the total number of $\BB$ events in the data sample assuming
equal production of $\Bz\Bzb$ and $\Bp\Bm$ at the $\FourS$.
The result for the branching fraction is ${\cal B}\left(\Bz\to\KS\Kpm\pimp\right) = \kskpiBFal$,
where the first error is statistical and the second is systematic.

We find the systematic error to be due to uncertainties in the signal PDFs
(5.2\,\%), including possible data-MC differences in the signal PDF shapes
evaluated using the control sample of $\Bz\to\Dm\pip,\Dm\to\KS\pim$ decays; 
uncertainties in the background PDFs (2.5\,\%), including effects due to the
fixed values of some of the \qqbar\ PDF parameters (recall that the
parametrization used is validated with off-peak and MC samples and that the
most critical parameters are floated in the fit to data; the uncertainties are
evaluated by varying the fixed parameters) and due to the fixed content of the
histograms used to describe the \BB\ background PDFs;
potential fit biases, studied using ensembles of simulated experiments where
continuum events are drawn from the PDF shapes and signal and \BB\ background
events are randomly extracted from MC samples (1.1\,\%);
uncertainties in the efficiency due to tracking (0.8\,\%), \KS\ selection
(0.9\,\%), and particle identification (2.8\,\%);
and the error in the number of \BB\ events (1.1\,\%).
We assign two systematic uncertainties to account for the nonuniform Dalitz
plot structure of the signal, both of which are estimated from MC simulations
with different resonant contributions: 
uncertainty in the fraction of misreconstructed events (3.0\,\%) and
uncertainty in the correction due to vetoes (4.1\,\%).
Other sources of systematic uncertainty, including the fixed yields of
$\BB_1$ and $\BB_4$, are found to be negligible (recall that the fitted yields
of $\BB_2$ and $\BB_3$ are consistent with expectation).

\begin{figure}[!htb]
\center
\includegraphics[width=.99\columnwidth]{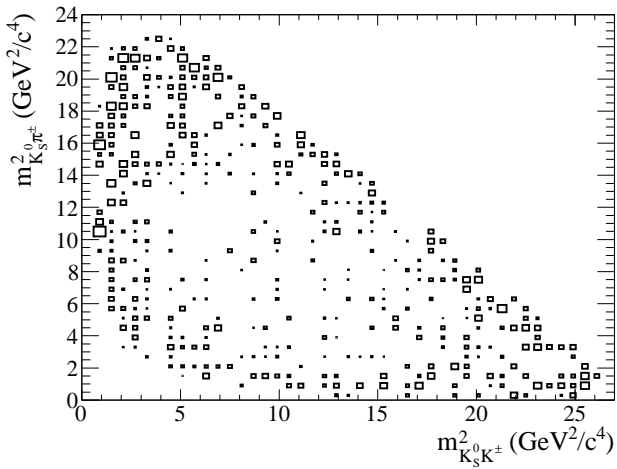}
\caption{
  Efficiency-corrected Dalitz plot distribution of $\Bz\to\KS\Kpm\pimp$ decays,
  obtained with the \splot\ technique~\cite{Pivk:2004ty}. 
   Bins with negative content appear empty as do regions corresponding to the charm
   and charmonium vetoes.
  The area of each box is proportional to the number of weighted events in
  that bin; the largest bin corresponds to 12 events.
}
\label{fig:signal-dalitz-splot}
\end{figure}

In Fig.~\ref{fig:signal-dalitz-splot} we show the
efficiency-corrected Dalitz plot for signal decays, obtained
using event-by-event signal probabilities.
We verify that this technique correctly reconstructs the
signal Dalitz plot distribution using MC simulations
in which the $\Bz\to\KS\Kpm\pimp$ events contain different structures.
There appears to be some structure in the $\Kstarz$ region at low
$\Kpm\pimp$ invariant mass, and an excess of events at low $\KS\Kpm$ mass
with a highly asymmetric helicity angle distribution.
Quantitative statements concerning the content of the Dalitz plot 
require a dedicated amplitude analysis, which is beyond the scope of the present study.
However it appears that there is no major contribution from an isospin partner
of the $f_{\rm X}(1500)$ decaying to $\KS\Kp$, which contrasts to the clear signal
seen in $\Bp\to\Kp\Km\pip$ decays~\cite{Aubert:2007xb}.

In summary, using the full \babar\ data sample of \onreslumi\ collected at the $\FourS$
resonance, we find evidence for charmless hadronic decays of neutral $\B$
mesons to the final state $\KS\Kpm\pimp$.
The signal has a significance of \nsigSys, after taking systematic effects
into account.  
We measure the branching fraction to be 
${\cal B}\left(\Bz\to\KS\Kpm\pimp\right) = \kskpiBFal$.
We convert this result to ${\cal B}[\Bz\to (\Kz\Kp\pim +\Kzb\Km\pip )]
= (6.4\pm 1.0\pm 0.6)\times 10^{-6}$ by multiplying by a factor of $2$.

We are grateful for the excellent luminosity and machine conditions
provided by our \pep2\ colleagues, 
and for the substantial dedicated effort from
the computing organizations that support \babar.
The collaborating institutions wish to thank 
SLAC for its support and kind hospitality. 
This work is supported by
DOE
and NSF (USA),
NSERC (Canada),
CEA and
CNRS-IN2P3
(France),
BMBF and DFG
(Germany),
INFN (Italy),
FOM (The Netherlands),
NFR (Norway),
MES (Russia),
MEC (Spain), and
STFC (United Kingdom). 
Individuals have received support from the
Marie Curie EIF (European Union) and
the A.~P.~Sloan Foundation.


\begin{thebibliography}{99}

\bibitem{Aubert:2006gm}
  B.~Aubert {\it et al.} (\babar\ Collaboration),
  Phys.\ Rev.\ Lett.\ {\bf 97}, 171805 (2006). 

\bibitem{Abe:2006xs}
  S.~W.~Lin {\it et al.} (Belle Collaboration),
  Phys.\ Rev.\ Lett.\ {\bf 98}, 181804 (2007).

\bibitem{Aubert:2007xc}
  B.~Aubert {\it et al.} (\babar\ Collaboration),
  Phys.\ Rev.\ Lett.\ {\bf 100}, 081801 (2008).

\bibitem{Aubert:2009vc}
  B.~Aubert {\it et al.} (\babar\ Collaboration),
  Phys.\ Rev.\ D {\bf 79}, 051102 (2009).

\bibitem{Chiang:2010ga}
  C.~C.~Chiang {\it et al.} (Belle Collaboration),
  Phys.\ Rev.\ D {\bf 81}, 071101 (2010).

\bibitem{Aubert:2006wu}
  B.~Aubert {\it et al.} (\babar\ Collaboration),
  Phys.\ Rev.\ D {\bf 74}, 072008 (2006).

\bibitem{Aubert:2007ua}
  B.~Aubert {\it et al.} (\babar\ Collaboration),
  Phys.\ Rev.\ D {\bf 76}, 071103 (2007).

\bibitem{Aubert:2006fha}
  B.~Aubert {\it et al.} (\babar\ Collaboration),
  Phys.\ Rev.\ D {\bf 75}, 012008 (2007).

\bibitem{Aaltonen:2008hg}
  T.~Aaltonen {\it et al.} (CDF Collaboration),
  Phys.\ Rev.\ Lett.\ {\bf 103}, 031801 (2009).

\bibitem{:2008ap}
  B.~Aubert {\it et al.} (\babar\ Collaboration),
  Phys.\ Rev.\ D {\bf 78}, 051103 (2008).

\bibitem{Aubert:2007fm}
  B.~Aubert {\it et al.} (\babar\ Collaboration),
  Phys.\ Rev.\ D {\bf 76}, 071104 (2007).

\bibitem{Aubert:2008rr}
  B.~Aubert {\it et al.} (\babar\ Collaboration),
  Phys.\ Rev.\ D {\bf 78}, 091102 (2008).

\bibitem{Aubert:2006aw}
  B.~Aubert {\it et al.} (\babar\ Collaboration),
  Phys.\ Rev.\ D {\bf 74}, 051104 (2006).

\bibitem{Aubert:2007xb}
  B.~Aubert {\it et al.} (\babar\ Collaboration),
  Phys.\ Rev.\ Lett.\ {\bf 99}, 221801 (2007).

\bibitem{Aubert:2008aw}
  B.~Aubert {\it et al.} (\babar\ Collaboration),
  Phys.\ Rev.\ D {\bf 79}, 051101 (2009).

\bibitem{Garmash:2003er}
  A.~Garmash {\it et. al.} (Belle Collaboration),
  Phys.\ Rev.\ D {\bf 69}, 012001 (2004).

\bibitem{Du:1995ff}
  D.~S.~Du and L.~B.~Guo,
  Z.\ Phys.\ C {\bf 75}, 9 (1997).

\bibitem{Ali:1998eb}
  A.~Ali, G.~Kramer, and C.~D.~Lu, Phys.\ Rev.\ D {\bf 58}, 094009 (1998).

\bibitem{Chen:1999nxa}
  Y.~H.~Chen {\it et al.},
  Phys.\ Rev.\ D {\bf 60}, 094014 (1999).

\bibitem{Deshpande:1997rr}
  N.~G.~Deshpande, B.~Dutta, and S.~Oh, Phys.\ Lett.\ B {\bf 473}, 141 (2000).

\bibitem{Du:2002up}
  D.~S.~Du {\it et al.},
  Phys.\ Rev.\ D {\bf 65}, 094025 (2002); {\bf 66}, 079904(E) (2002).

\bibitem{Beneke:2003zv}
  M.~Beneke and M.~Neubert,
  Nucl.\ Phys.\ B {\bf 675}, 333 (2003).

\bibitem{Chiang:2003pm}
  C.~W.~Chiang {\it et al.},
  Phys.\ Rev.\ D {\bf 69}, 034001 (2004).

\bibitem{Guo:2006uq}
  L.~Guo, Q.~G.~Xu, and Z.~J.~Xiao,
  Phys.\ Rev.\ D {\bf 75}, 014019 (2007).

\bibitem{Wang:2006sy}
  R.~M.~Wang {\it et al.},
  Eur.\ Phys.\ J.\ C {\bf 47}, 815 (2006).

\bibitem{Aubert:2001tu}
  B.~Aubert {\it et al.} (\babar\ Collaboration),
  Nucl.\ Instrum.\ Methods\ Phys.\ Res., Sect.\ A {\bf 479}, 1 (2002).

\bibitem{Amsler:2008zz}
  C.~Amsler {\it et al.} (Particle Data Group),
  Phys.\ Lett.\ B {\bf 667}, 1 (2008).

\bibitem{Aubert:2009az}
  B.~Aubert {\it et al.} (\babar\ Collaboration),
  Phys.\ Rev.\ D {\bf 79}, 072006 (2009).

\bibitem{Aubert:2007sd}
  B.~Aubert {\it et al.} (\babar\ Collaboration),
  Phys.\ Rev.\ Lett.\ {\bf 99}, 161802 (2007).

\bibitem{Aubert:2007vi}
  B.~Aubert {\it et al.} (\babar\ Collaboration),
  Phys.\ Rev.\ D {\bf 80}, 112001 (2009).

\bibitem{hfag}
  E.~Barberio {\it et al.} (Heavy Flavor Averaging Group),
  arXiv:0808.1297, and online update at
  {\tt http://www.slac.stanford.edu/xorg/hfag/}.

\bibitem{Albrecht:1990am}
  H.~Albrecht {\it et al.} (ARGUS Collaboration),
  Phys.\ Lett.\ B {\bf 241}, 278 (1990).

\bibitem{Pivk:2004ty}
  M.~Pivk and F.~R.~Le Diberder,
  Nucl.\ Instrum.\ Methods\ Phys.\ Res., Sect.\ A {\bf 555}, 356 (2005).

\end{thebibliography}
\end{document}